\documentstyle[preprint,prl,aps,epsf]{revtex}

\newcommand{\up}{\uparrow}
\newcommand{\down}{\downarrow}

\renewcommand{\[}{\begin{equation}}
\renewcommand{\]}{\end{equation}}

\begin{document}
\draft
\title{Parallelization of the exact diagonalization of the $t-t'$-Hubbard model}
\author{W. Fettes, I. Morgenstern, T.\ Husslein}
\address{Universit{\"a}t Regensburg, Fakult{\"a}t Physik, 93040 Regensburg, Germany\\
email: Werner.Fettes@physik.uni-regensburg.de}
\date{29 August 1997}

\maketitle

\begin{abstract}
We present a new parallel algorithm for the exact diagonalization
of the $t-t'$-Hubbard model with the Lanczos-method.
By invoking a new scheme of labeling the states we were able to obtain
a speedup of up to four on 16 nodes of an IBM SP2 for the calculation of
the ground state energy and an almost linear speedup for the calculation of
the correlation functions.
Using this algorithm we performed an extensive study of the influence of
the next-nearest
hopping parameter $t'$ in the $t-t'$-Hubbard model on ground state energy and
the superconducting correlation functions for both attractive
and repulsive interaction.
\end{abstract}

\section{Introduction}

The Hubbard model \cite{HUB63} is one of the generic models in 
many particle physics.
Due to difficulties with the analytic
solution of the Hubbard model in two dimensions, this model is intensively
studied with various numerical algorithms,
e.g.\ exact diagonalization \cite{DAG90},
\cite{LEU92}, \cite{LIN93} \cite{FET95}, stochastic diagonalization \cite{RAE92}, \cite{MIC96} and
quantum Monte Carlo algorithms \cite{SOR89}, \cite{MOR93},
\cite{LIN92}, \cite{HUS94}.

The single band Hubbard model with
additional next nearest neighbor hopping is given in real the space
by:
\[
\label{eqhubbard}
H =
-t \sum_{<i,j>,\sigma} c_{i,\sigma}^\dagger c^{}_{j,\sigma}
-t' \sum_{<<i,j>>,\sigma} c_{i,\sigma}^\dagger c^{}_{j,\sigma}
+ U \sum_i n_{i,\uparrow} n_{i,\downarrow}
\quad.
\]
The sum $<i,j>$ is over the nearest neighbors and $<<i,j>>$ is the
sum over the next nearest neighbors. $c_{i,\sigma}^\dagger$ is the
creation operator for an electron with spin $\sigma$ on site $i$ and
$n_{i,\sigma}$ is the corresponding number operator. 
Throughout this article we take $t=1$ as energy unit.

In the momentum space this Hamiltonian reads as:
\[
H = \sum_{k,\sigma} \varepsilon_{k} c_{k,\sigma}^\dagger c^{}_{k,\sigma} +
U \sum_{k,p,q} c_{k,\uparrow}^\dagger c_{p,\downarrow}^\dagger
c^{}_{p-q,\downarrow} c^{}_{k+q,\uparrow}
\]
with
\[
\label{eqepsilon}
\varepsilon_k = -2t \big( \cos(k_x) + \cos(k_y) \big)
-4 t' \cos(k_x) \cdot \cos(k_y) 
\quad .
\]

The usual Hubbard-model ($t'=0$) has a Van Hove singularity in the density of states
at half filling in the noninteracting case ($U=0$).
It is  possible to move this Van Hove singularity to
any electron filling by extending the Hubbard model
by an additional next nearest neighbor hopping $t'$-term in the kinetic energy.
The $t-t'$-Hubbard
model \cite{NEWN91} shows superconductivity for repulsive interactions
$U$ with d$_{x^2-y^2}$-symmetry \cite{HUS94} and for attractive interaction
with on site s-symmetry \cite{SCA89}, \cite{LOH89}, \cite{BOR92}.

As a measure for the superconductivity we calculate the reduced
two particle density matrix according to the concept of Yang
\cite{YAN62}. From this two particle density matrix we calculate the
two particle correlation functions for different symmetries and additionally
we calculate the vertex correlation function \cite{WHI89}.

The Van Hove scenario predicts an increase of $T_c$ for fillings close to a Van Hove
singularity \cite{NEWN92}.
We study the influence of the Van Hove singularity on the
superconducting correlation functions by modifying
the $t'$-hopping parameter for fixed fillings.
Here we use
the Lanczos-algorithm \cite{CUL85}, \cite{LEU92}
as exact diagonalization technique to determine the
ground state and from there ground state
properties of the $t-t'$-Hubbard model.

The basic limitations on the calculations of large system sizes with
the  Lanczos-algorithm is the huge memory consumption of this method.
But to our surprise we found that even for Hubbard systems of size
$4\times 4$, which the Lanczos method is capable of handling, the CPU consumption
of the simulations was substantial and
made a detailed scan of the parameter space given by interaction strength $U$,
filling $n$ and next nearest neighbor hopping $t'$ almost impossible.
We therefore implemented the
Lanczos-method on IBM SP2 parallel computer with MPI (Message Passing Interface)
for the communication between the processes to speed up the  calculations.

\section{Exact Diagonalization with the Lanczos-Algorithm}

Before turning our attention to the parallel techniques we
outline the basic concept of the Lanczos method \cite{CUL85}.
In the case of the exact diagonalization the
Hamiltonian $H$ of the system is written in matrix or Heisenberg
representation.
One chooses an orthonormal single particle basis,
to represent the
many-particle states. For the $t-t'$-Hubbard model we use the
momentum-space representation.
Each many particle basis state $| \Phi_i \rangle$ is
a product of the spin up $|\Phi_{i_{\uparrow},\uparrow} \rangle$ and the
spin down $| \Phi_{i_{\downarrow},\downarrow}\rangle $ component:
\[
\label{EqPhi}
|\Phi_i\rangle \equiv | \Phi_{i_{\uparrow},\uparrow} \rangle
\otimes | \Phi_{i_{\downarrow},\downarrow} \rangle \equiv
c_{k_1,\up}^\dagger c_{k_2,\up}^\dagger \ldots c_{k_{n_\up},\up}^\dagger
\cdot
c_{p_1,\down}^\dagger c_{p_2,\down}^\dagger \ldots
c_{p_{n_\down},\down}^\dagger
| 0 \rangle \quad ,
\]
where $c_{k_i,\sigma}^\dagger$ is the creation operator of an electron with
momentum $k_i$ and spin $\sigma$. $| 0 \rangle $ is the vacuum state.

Each many particle state $| \Psi \rangle$ can be represented by means of 
the basis states $|\Phi_i\rangle$ and coefficients $\alpha_i$:
\[
\label{eqpsi}
| \Psi \rangle = \sum_{i=1}^{M} \alpha_i | \Phi_i \rangle
\quad .
\]

The Hilbert space of a system consisting of a lattice of  $L$ sites and $n_\uparrow$
electrons with spin up and $n_\downarrow$ electrons with spin down has the
dimension
\[
M = M_\up \cdot M_\down = {L \choose n_\up} \cdot {L \choose n_\down}
\quad .
\]
As the size of the Hilbert space also determines the size of the computer memory,
that is used in the calculation,
one tries to reduce the Hilbert space.
The usual way to restrict the size of the Hilbert space $M$ is to apply
symmetries of the lattice and the Hamiltonian. The translation invariance is a
symmetry  easily  implemented when solving the $t-t'$-Hubbard model in momentum
space representation with the Lanczos algorithm. The Hilbert space
decomposes into subspaces $M^K_t$ containing only
basis states $| \Phi_i \rangle$, which have the
same total momentum $K$:
\[
\label{EqK}
K = K_\up + K_\down =
\sum_{i=1}^{n_\up} k_i + \sum_{j=1}^{n_\down} p_j
\quad ,
\]
where $k_i$ and $p_j$ are the momenta of the creation operators
$c_{k_i,\uparrow}^\dagger$ resp.\ $c_{p_j,\downarrow}^\dagger$
in eq.\ \ref{EqPhi}.
We denote the number of states of particles with the same spin with the total 
momentum $K_\up$ as $M^\sigma_{K_\uparrow}$. 
In a $4\times 4$ system with $n_\sigma = 4$ the number of states $M^\sigma_k$ 
varies for different momentum between $112$ and $120$.
The states $| \Phi_{i_\downarrow,\downarrow}
\rangle$ must have the total momentum $K_\downarrow = K - K_\uparrow$, so that
the product state $|\Phi_{i_\uparrow,\uparrow} \rangle \otimes
|\Phi_{i_\downarrow,\downarrow} \rangle$ is in the subspace $M^K_t$.
Altogether the subspace $M^K_t$ has the size
\[
M^K_t = \sum_{K_\uparrow=1}^L M^\uparrow_{K_\uparrow} \cdot 
M^\downarrow_{K-K_\uparrow}
\quad .
\]
For various fillings $n_\up = n_\down$ 
table \ref{tabstates} gives an overview.

When storing the coefficients $\alpha_i$ with 8 Byte floating point numbers
the memory demand for one state is $8\cdot M =8 \cdot 4368^2 \approx 146$ MByte 
for a lattice with 16 sites and $n_\up = n_\down = 5$ electrons.
Even if one uses the translation 
symmetry it remains a demand of $8 \cdot M_t^K  \approx 9 $ MByte for each state.
In the standard Lanczos--algorithm \cite{CUL85}, \cite{GOL89} it is
necessary to store three many particle states $|\Psi\rangle$ 
to calculate the ground state and the ground state energy.

We implemented the Lanczos-iteration \cite{GOL89}:
\[
\label{eqlanczos}
\gamma_j = \langle \Psi_j | \Big( H \cdot | \Psi_j  \rangle - \beta_{j-1}
| \Psi_{j-1} \rangle \Big)
\quad ,
\]
with the coefficients $\gamma_j$ (diagonal) and $\beta_j$
(offdiagonal) of the tridiagonal
matrix $T_j$ and the Lanczos-vectors $|\Psi_j \rangle$.
From this tridiagonal matrix $T_j$ we calculate the eigenvalues of $H$.
In the Lanczos-scheme it is not necessary to transform the
matrix $H$. Therefore it is even not necessary to store the
matrix elements $\langle \Phi_i | H | \Phi_j \rangle$; they are only calculated,
when they are needed for the further evaluation of the Lanczos-iteration
(eq.\ \ref{eqlanczos}).

\section{Parallelization of the Lanczos--Algorithm}

In this section we concentrate our effort on how to  speed up
the simulations with the Lanczos method.
The determination of the ground state properties of a Hamiltonian with the
Lanczos--algorithm consists of two main parts concerning the
consumption of CPU-time. First the ground state energy $E_0$ and the
ground state $|\Psi_0 \rangle$ are calculated.
The second main part is the determination of the two--particle
density matrix \cite{YAN62}:
\[
\label{eqvarrho}
\renewcommand{\arraystretch}{2.0}
\begin{array}{rl}
\varrho_{k_1,k_2,k_3,k_4} & \equiv
\langle
c^\dagger_{k_1,\up} c^\dagger_{k_2,\down} c^{}_{k_3,\down} c^{}_{k_4,\up}
\rangle \\
& =
\sum\limits_{i,j} \alpha_i \alpha_j \langle \Phi_i |
c^\dagger_{k_1,\up} c^\dagger_{k_2,\down} c^{}_{k_3,\down} c^{}_{k_4,\up}
| \Phi_j \rangle
\end{array}
\]
as the main observable of interest.

\subsection{Algorithms for handling the basis states}

To handle an arbitrary state $|\Psi\rangle$ 
it is necessary to know for each basis state $|\Phi_i\rangle$ the 
indices $k_i$ and $p_i$ of the creation operators in eq.\ \ref{EqPhi} 
and the weights $\alpha_i$. 

A simple possibility for such an algorithm is the bitcoding or 
bitrepresentation of the basis states.
Here the momenta $k_i$ are labeled from $1$ to $L$ and the momenta $p_i$ from 
$L+1$ to $2L$. Then the states $|\Phi_i\rangle$ are expressed by an one dimensional array of bits, which 
are one for occupied sites, and otherwise the bits are zero. 
If one interprets this array as binary 
representation of an integer number one has an algorithm to assign each basis state
$|\Phi_j\rangle$ an index $j$. As example we take a lattice with 4 sites and each two 
electrons with spin up and down:
\[
|\Phi_j \rangle = 
c^\dagger_{1,\uparrow} c^\dagger_{3,\uparrow} \cdot
c^\dagger_{2,\downarrow} c^\dagger_{3,\downarrow} |0\rangle 
\leftrightarrow
1010\,0110
\leftrightarrow
 166 = j
\leftrightarrow
\alpha_j
\quad .
\] 
For any many particle state only the coefficients $\alpha_j$ need to be stored.

But the bitrepresentation has the great disadvantage, that a
huge amount of memory is wasted, because the bitrepresentation of many integer numbers $j$ does
not correspond to a valid basis state.
For a system with $L$ latticepoints in an array of the length $(2^L)^2$ 
one only stores $M \ll (2^L)^2$ numbers. 
For the above example $L= 16$
and $n_\up = n_\down = 5$ there is $(2^{16})^2 / M_t^K  \approx 3602$.

Therefore it is desirable to use another algorithm 
($|\Phi_j\rangle \leftrightarrow j$).
One example is the hashing algorithm \cite{GAG86}.

We developed a new algorithm. 
Though we only present results for the  momentum  space representation,
this algorithm can also be implemented very efficiently 
for the exact diagonalization in real space \cite{FET94}.

\subsection{\label{numbering}Numbering of the states}

First we are numbering the momenta from 1 to $L$ ($k_i \in \{
1,\ldots,L\}$).
Second we define $k_1 < k_2 < \ldots < k_{n_\sigma}$ to fix the
sign.

The state with the number $1$ is 
\[
| \Phi_{1,\sigma} \rangle =
c_{1,\sigma}^\dagger
c_{2,\sigma}^\dagger \ldots
c_{{n_\sigma},\sigma}^\dagger
| 0 \rangle
\quad .
\]
In $| \Phi_{2,\sigma} \rangle $ the electron $c_{{n_\sigma},\sigma}^\dagger$
moves from $n_\sigma$ to $n_\sigma+1$. In 
$| \Phi_{L-{n_\sigma}+1,\sigma} \rangle $ this "last" electron has reached the
latticepoint $L$. Next in $|\Phi_{L-n_\sigma +2}\rangle$ the two
creation operators with the highest index are increased by one,
\[
|\Phi_{L-n_\sigma+2,\sigma}\rangle = 
c^\dagger_{1,\sigma} c^\dagger_{2,\sigma} \ldots c^\dagger_{n_\sigma-2,\sigma}
c^\dagger_{n_\sigma,\sigma} c^\dagger_{n_\sigma+1,\sigma} |0\rangle
\quad .
\]
Then the "last" creation operator moves. 
These are the states with the numbers $L-n_\sigma+2$ to $2L-2n_\sigma+1$.
Next the "last" two electrons go to the sites $n_\sigma+1$
and $n_\sigma+2$. If both electrons have reached the final two lattice sites
($L-1$, $L$) three electrons move in the same manner through the
lattice.

One gets the number $i_\sigma$ of the state 
$|\Phi_{i_\sigma,\sigma} \rangle$ with
\[
i_\sigma = \sum_{m=1}^{n_\sigma} \left(
\sum_{j=l_{m-1}+1}^{l_m-1} {L-j \choose n_\sigma -m}
\right)
\quad ,
\]
where $l_m$ is the position of the electron $m$ in the lattice.

Using the translational invariance of the Hamiltonian in the k-space
representation means, that we keep the total momentum $K$ 
of the basis states $| \Phi_i \rangle $ fixed.
In this case one can only choose the "spin-up" part of the basis state free
and take such a "spin-down" state, that eq.\ \ref{EqK} is full filled.
This means, we must use another convention to label the states.

We generate all states
$|\Phi_{i,\sigma}\rangle$ in the sequence as described above and calculate the momentum.
For each momentum we count independently the indices.

The coefficients $\alpha_i$ are now labeled in following way:
First we combine with state number 1, momentum 1 and spin up
with all possible states with spin down for a given $K$.
Then we do the same with state number 2, momentum 1 and spin up.
Next we switch to state 1 with momentum 2 and spin up and so on.

\subsection{Dividing up the memory}

In the parallel algorithm each of the $P$ processes
stores the coefficients $\alpha_i$ of $M_P \equiv M^K_t/P$
basis states.
$\alpha_i$ is stored on process $p = i/M_P$.

\subsection{\label{sectionmatmult}Matrix-vector multiplication}

In the Lanczos method it is necessary to perform a matrix-vector multiplication
between the matrix $H$ and a Lanczos-vector $|\Psi_j\rangle$.
In the parallel algorithm this is carried out in the following way:
For $i=1,\ldots M_P$:
\begin{itemize}
\item Each process $p$ calculates the index numbers $j$ of the basis states
$H \cdot |\Phi_{i+(p-1)M_P}\rangle$, the multiplication results
for the $i$-th state and the process, on which the states $j$ are stored.

\item Each process exchanges the multiplication results
and the numbers $j$ of the
basis states, which are not stored on the process, with 
process $p_j = j / M_P$.

\end{itemize}

\subsection{\label{sectionvarrho}Calculation of the reduced two particle density matrix}

To calculate the two-particle density matrix $\varrho_{k_1,k_2,k_3,k_4}$ in
an efficient way, one only calculates the elements
\[
\langle \Phi_m |
c_{k_1,\up}^\dagger c_{k_2,\down}^\dagger c^{}_{k_3, \down} c^{}_{k_4,\up}
| \Phi_n \rangle
\quad ,
\]
which are nonzero. That means one takes a basis state
$| \Phi_n \rangle$, one of the $L^4$ possible combinations of
$k_1$, $k_2$, $k_3$ and $k_4$ and applies
$c_{k_1,\up}^\dagger c_{k_2,\down}^\dagger c^{}_{k_3, \down} c^{}_{k_4,\up}$ 
to $|\Phi_n\rangle$,
afterwards one calculates the index number of this transformed basis state
$| \Phi_m \rangle$.

Each process stores the complete matrix $\varrho_{k_1,k_2,k_3,k_4}$.
This matrix takes for example in the $4\times 4$ system $8\cdot 16^4$ Byte or $512$
KByte of memory.

In this algorithm one has to calculate $N_1 = M_t^K \cdot L^4$ expectation values,
compared to $N_2 = ({M_t^K})^2$ expectation values that would be  calculated if one took all
combinations $| \Phi_i \rangle $ and $| \Phi_j \rangle$ into account.
In a $4\times4$ system with $n_\up = n_\down = 5$ electrons 
$N_2 / N_1 \approx 18$ and the amount of saved CPU-time is significant. 
But in a $4\times 4$ system with only $n_\up = n_\down = 3$ electrons 
$N_2/N_1 \approx 0.3$ and this algorithm is slower.

To calculate $\varrho_{k_1,k_2,k_3,k_4}$
we use a similar way for the exchange of the weights and numbers as for
matrix-vector multiplication (see section \ref{sectionmatmult}).

At the end the values of the arrays $\varrho_{k_1,k_2,k_3,k_4}$ of
all processes are summed on one process.

\section{Performance Analysis of the parallel code}

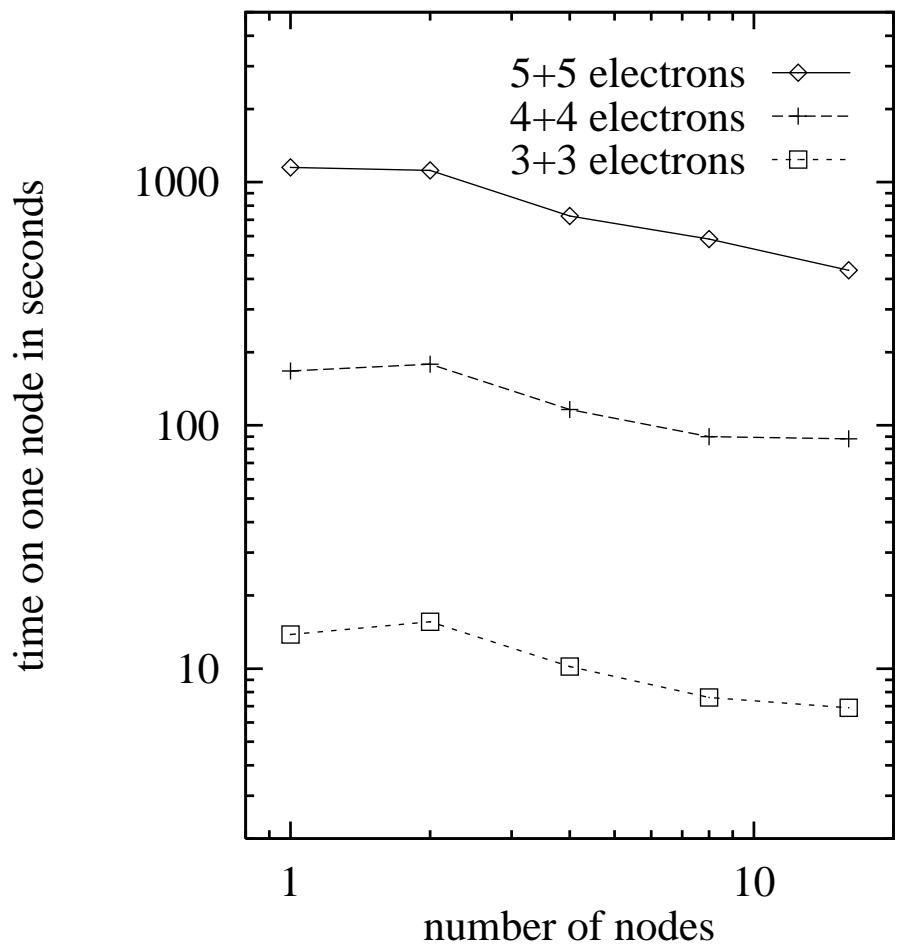
\begin{figure}[p]
\caption{
\label{cpuiter}
CPU-time for one Lanczos-iteration in a $4\times4$ lattice.}
\end{figure}

First we take a look on the dependence of the CPU time of one
Lanczos iteration (eq.\,\ref{eqlanczos}) for a different number of processes. As example
we use a $4\times4$ lattice with three different numbers of electrons
(\,fig.\ \ref{cpuiter}\,).

The small decay from 1 to 2 processes is
due to communication between processes which is only necessary
for more than one process.
Then one sees as expected a decrease of computation time.
As expected this decrease vanishes for an increasing  number of processes,
since the communication is growing with the number of processes.

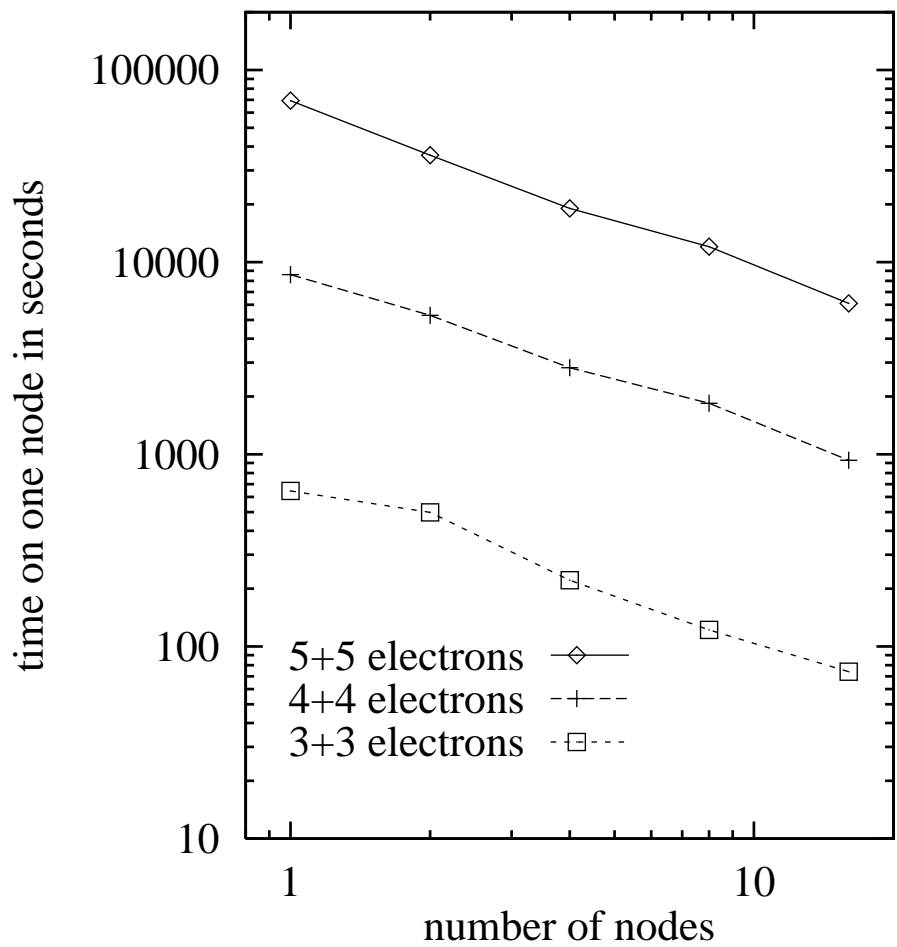
\begin{figure}[p]
\caption{
\label{cpucorr}
CPU-time for the determination of the two-particle
density matrix $\varrho_{k_1,k_2,k_3,k_4}$ in a $4\times4$ lattice.}
\end{figure}

In figure \ref{cpucorr} we examine the  CPU-consumption for the two-particle 
density matrix
for the same system size and fillings as in figure \ref{cpuiter}.
Here the gain of time is much larger than for the calculation of the
energy. In this case  more computation is performed for the determination of
one matrix element. 
For more than 2 processes the dependence of nodes and CPU-time
is nearly linear.

This can be understood, if we look on the numbering of the states.
Most of the states $H \cdot | \Phi_i \rangle$ are stored on the
same node and only for a fraction of these states the weights
must be interchanged with an other process.

Summarizing the algorithm for this parallel implementation of
the Lanczos-iteration and the determination of the two-particle
density matrix is a coarse grained algorithm and therefore
achieves a good speed-up on a
parallel computer like
the IBM SP2, with only some, but very powerful, processors.
For very many processors the communication grows dramatically and
no further speed-up is reachable.

\section{\label{secresults}Numerical Results}

Now we want to turn our attention from the technical points of view to
physical properties of the $t-t'$-Hubbard model.
Especially we study the influence of the  next nearest
hopping parameter $t'$ on the ground state energy and superconducting 
correlation functions in the ground state of a $4\times 4$ cluster.
We focus to $n_e\equiv n_\uparrow = n_\downarrow = 5$ electrons which corresponds
to a filling $\langle n \rangle = 0.625$. This is a so called 
closed shell situation, which can be also handled with the
projector quantum Monte Carlo method \cite{BOR92}.

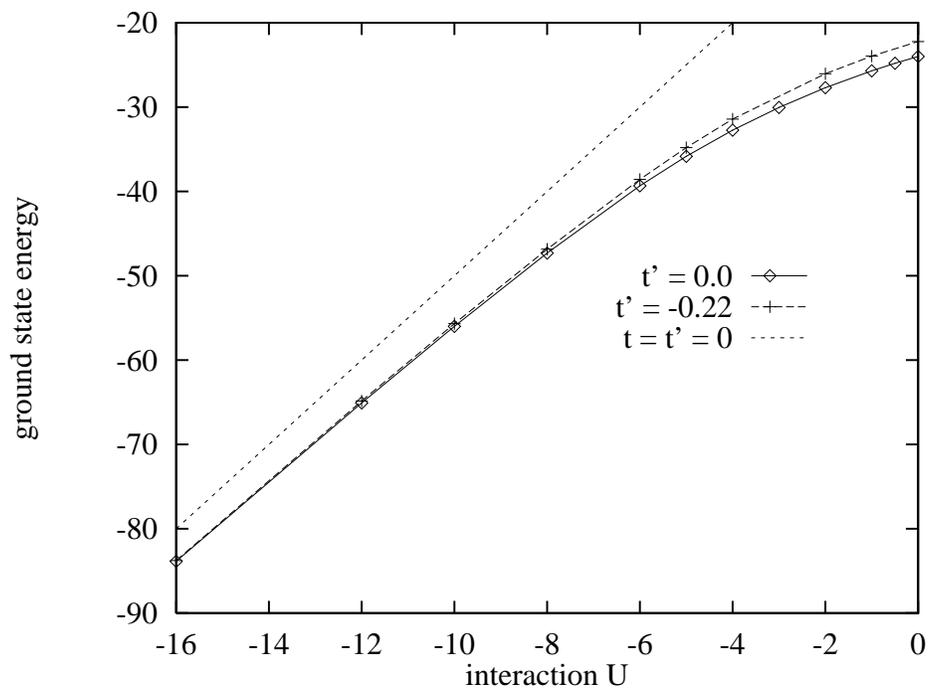
\begin{figure}[p]
\caption{
\label{emin}
Ground state energy in a $4\times4$ lattice with $n_\up = n_\down = 5$
electrons for the $t-t'$-Hubbard model.}
\end{figure}

First we study the ground state energy $E_0$ in the attractive
$t-t'$-Hubbard model (\,fig.\ \ref{emin}\,). For small and intermediate interaction
strength ($|U| \le 10$) there is a visible difference between
the energy with $t'=0$ and $t'=-0.22$. This difference results from
the changes in the structure of $\varepsilon_k$ (eq.\,\ref{eqepsilon})
with $t'$. But for large interaction strength $|U|$ the influence 
of the kinetic part is vanishing.
In this interaction regime the ground state energy is approximately
linear with $U$ and approaches slowly the energy $Un_e$ of the system without 
hopping. 

Next we turn our attention to the superconducting correlation 
functions.
In the concept of off diagonal long range order \cite{YAN62}
the largest eigenvalue and eigenvector of 
$\varrho_{k_1,k_2,k_3,k_4}$ is calculated. 
The quantum Monte Carlo algorithms handle system sizes,
where it is impossible to calculate the complete two-particle density matrix
due to the memory consumptions \cite{FET96}.
In order to compare the exact diagonalization results with the  
quantum Monte Carlo
data we study  two-particle correlation functions for certain 
symmetries \cite{MOR94}, e.g.
\[
\renewcommand{\arraystretch}{2.0}
\begin{array}{rcl}
C_{\rm s} (r) & = & \frac{1}{L} \sum\limits_{j} \langle c_{j,\up}^\dagger c_{j,\down}^\dagger
c^{}_{j+r,\down}  c^{}_{j+r,\up} \rangle \quad ,\\
C_{\rm d} (r) & = & \frac{1}{L} \sum\limits_{j} \sum\limits_{\delta,\delta'}
g_\delta g_{\delta'}
\langle c_{j,\up}^\dagger c_{j+\delta,\down}^\dagger
c^{}_{j+r+\delta',\down}  c^{}_{j+r,\up} \rangle \quad ,\\
\end{array}
\]
where the index s denotes the on site s-wave symmetry and
d the $d_{x^2-y^2}$-wave symmetry.
The factor $g_\delta = \pm 1$ gives the signs of the d-wave.
($+1$ in x- and $-1$ in y-direction)
These full correlation functions have  nonzero values 
even for a system with no interaction.
Responsible for this are the one-particle correlation functions
\[
C_{\rm o}^\sigma (r)
 = \frac{1}{L} \sum\limits_{j} \langle c_{j,\sigma}^\dagger
c^{}_{j+r,\sigma}  \rangle
\quad ,
\]
which decay to zero with
$\frac{1}{|r|}$ and thus do not really contribute to the 
long range behavior that signals superconductivity.
To exclude the contribution of the one-particle
correlation functions we define the vertex correlation function 
as
\[
\renewcommand{\arraystretch}{2.0}
\begin{array}{rcl}
C_{\rm s}^{\rm v} (r) & = & C_{\rm s} (r)  -
C_{\rm o}^\up (r) \cdot C_{\rm o}^\down (r) \quad , \\
C_{\rm d}^{\rm v} (r) & = &
C_{\rm d} (r)  - \sum\limits_{\delta,\delta'} \left( g_\delta
g_{\delta'}
C_{\rm o}^\up (r) \cdot C_{\rm o}^\down (r+\delta-\delta')
\right) \quad .
\end{array}
\]

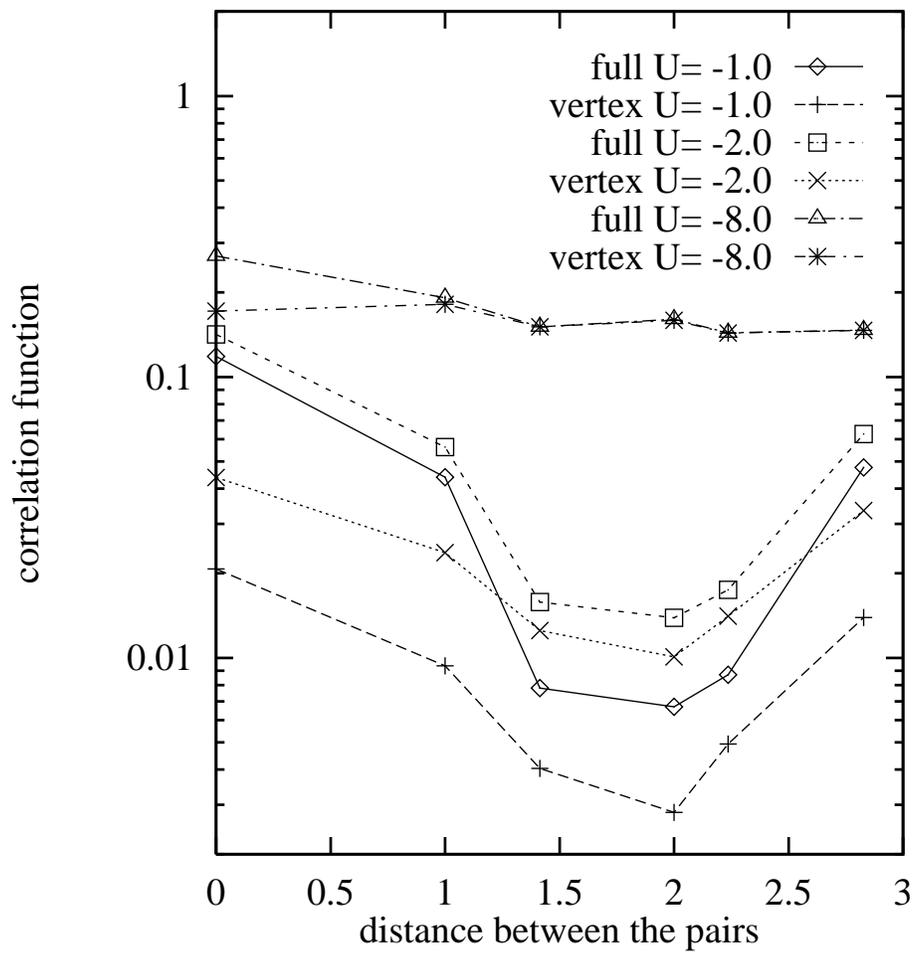
\begin{figure}[p]
\caption{
\label{corrvgl}
Full (${C}_{\rm s}(|r|)$) and vertex correlation 
(${C}_{\rm s}^{\rm v}(|r|)$)
function with on site s--wave symmetry in
a $4\times4$ lattice with $n_\up = n_\down = 5$ electrons in the
$t-t'$-Hubbard model with $t'=-0.22$.}
\end{figure}

In figure \ref{corrvgl} we show $C_{\rm s}$ and $C_{\rm s}^{{\rm v}}$ in
dependence of the distance $|r|$.
For $U=-8$ there is a visible difference only for $r=0$.
For $U=-1$ the one-particle contributions are dominant.
In this case it is important to study the vertex correlation function 
to get the "superconducting" correlations. But already for $U=-2$ and 
$|r| \ge \sqrt{2}$ the difference between full and vertex correlation function
is less than 30\% and it is less important to take $C^\sigma_{\rm o}$
in account.

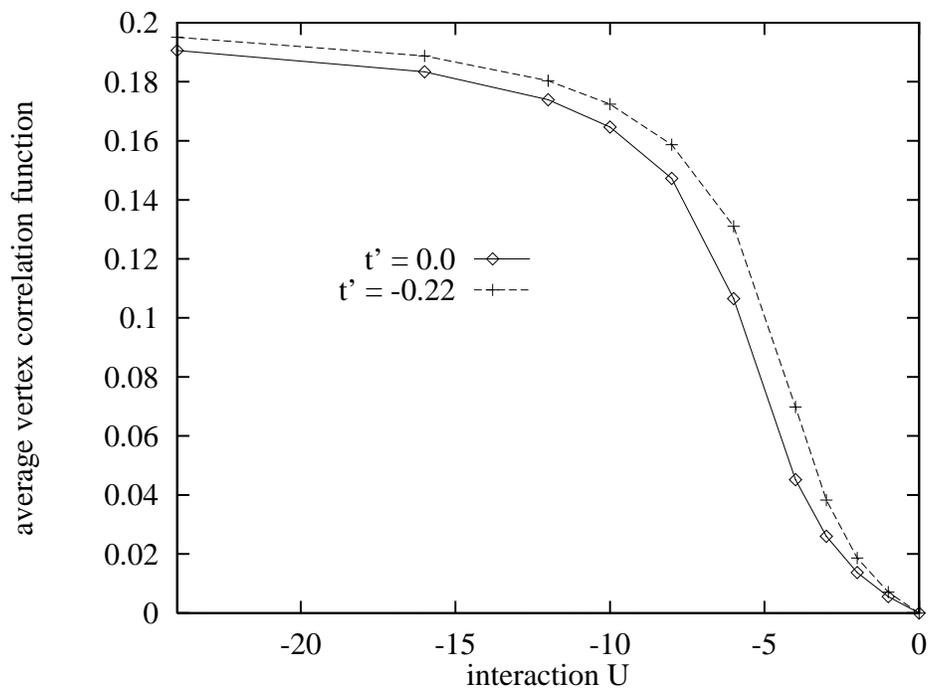
\begin{figure}[p]
\caption{
\label{neguvertex}
Average vertex correlation function ($\bar{C}_{\rm s}^{\rm v}$)
with on site s--wave symmetry in
a $4\times4$ lattice with $n_\up = n_\down = 5$ electrons in the
$t-t'$-Hubbard model.} 
\end{figure}

As a measure for the superconductivity in a system we show in 
fig.\ \ref{neguvertex} the average of the vertex correlation function
\[
\bar{C}_{\rm s}^{\rm v} = \frac{1}{L} \sum\limits_{i}
C_{\rm s}^{\rm v} (i)
\quad .
\]

For a small interaction strength $|U|$ the increase of the 
correlation functions is small.
Between $U=-2$ and $U=-10$ there is a strong increase.
Finally at $|U|>10$ the curves flatten. 

For correlation functions the
influence of the additional hopping $t'$ remains important even if there
is nearly no difference in the energies (cp.\ fig.\ \ref{emin} 
and \ref{neguvertex}, $U<-10$). 

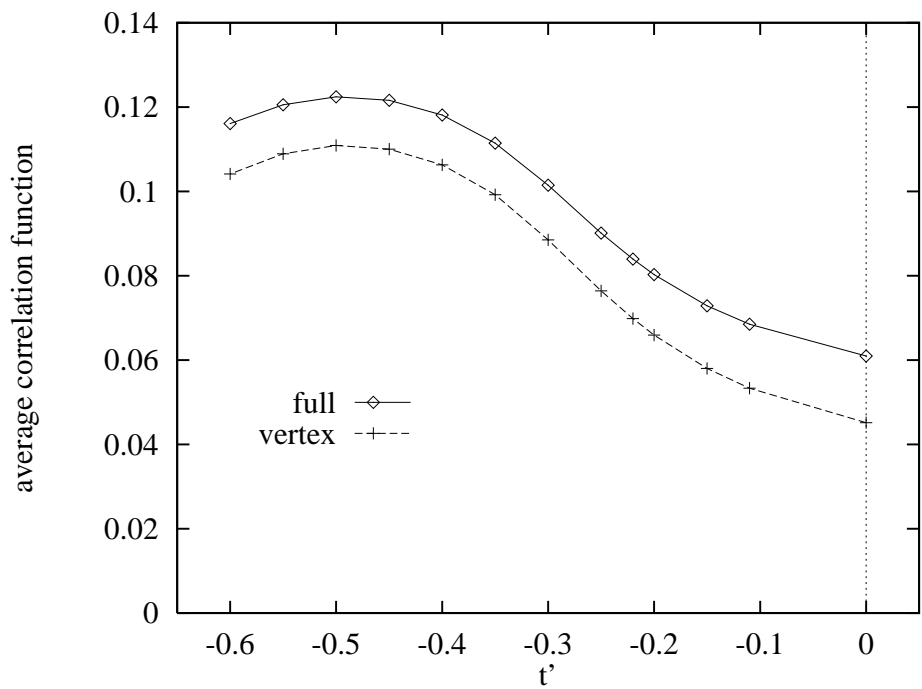
\begin{figure}[p]
\caption{
\label{negutprime}
Average full ($\bar{C}_{\rm s}$) and average vertex correlation
($\bar{C}_{\rm s}^{\rm v}$) function with on site s--wave symmetry in
a $4\times4$ lattice with $n_\up = n_\down = 5$ electrons in the
$t-t'$-Hubbard model and the attractive interaction $U=-4$.}
\end{figure}

Therefore we study the influence of $t'$
for the interaction $U=-4$ (\,fig.\ \ref{negutprime}\,). 
The correlation functions have a broad maximum around $t'=-0.50$. 
As it is commonly accepted \cite{BOR92} the
finite size gap has an influence on the superconducting correlation
functions. The finite size gap $g(t')$ in the
energy dispersion $\varepsilon_k$ (\,eq.\ \ref{eqepsilon}\,) 
of the free system ($U=0$) between the highest
occupied state and the lowest unoccupied state is given in this case by:
\[
g(t') =
\begin{cases}
{\begin{array}{llr}
2 - 4 \cdot |t'| & \quad \mbox{for} & -0.50 \le t' \le 0 \\
0  & \quad \mbox{for} &  t' \le -0.50 \\
\end{array}
\quad .
}
\end{cases}
\]
This means the finite size gap is becoming smaller with increasing $|t'|$. 
According to \cite{BOR92} this will lead to an increase in the 
superconducting correlations. Figure \ref{negutprime} confirms this
for $t'>-0.50$. But at $t'<-0.50$ the correlations decrease again. 
Therefore also the structure of the energy dispersion $\varepsilon_k$
has an influence on the correlation functions.

In figure \ref{negutprime} the maximum of the correlation functions is not
at $t'=-0.37$, where the noninteracting system has a Van Hove singularity in the thermodynamic limet.
There a maximum of $T_c$ is predicted by the Van Hove scenario \cite{NEWN92}. 
But for small system sizes one cannot really speak of a Van Hove singularity.
Therefore the results of figure \ref{negutprime} are not in contradiction to
the Van Hove scenario.

Yet, the question remains, whether the observed increase of the 
correlation function
results only from a vanishing finite size gap  or is related to changes of the
Fermi surface. To clarify this point it will be necessary to calculate larger
systems.

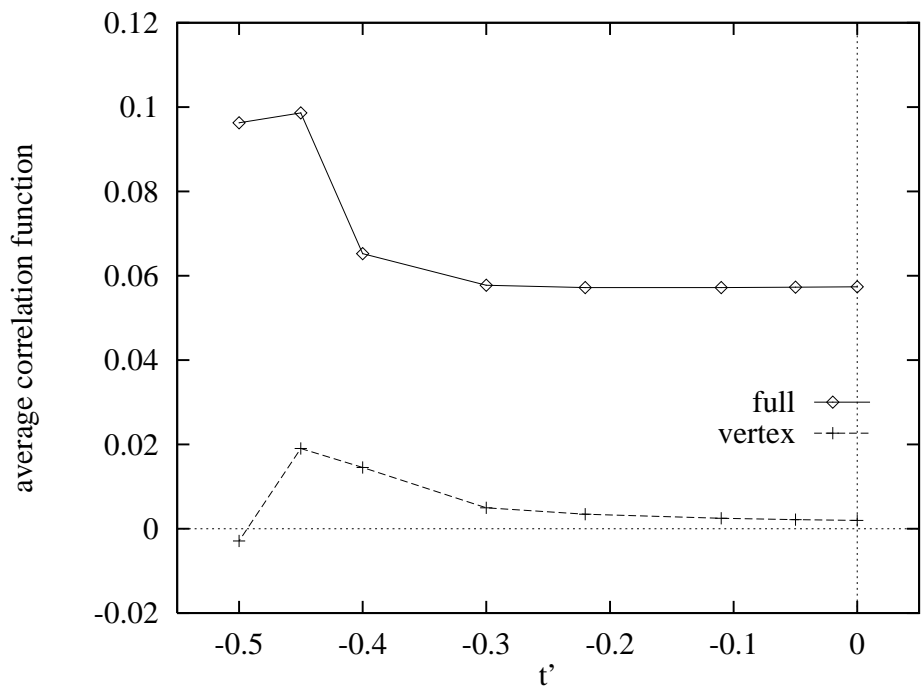
\begin{figure}[p]
\caption{
\label{posutprime}
Average full ($\bar{C}_{\rm d}$) and average vertex correlation
($\bar{C}_{\rm d}^{\rm v}$) function with d$_{x^2-y^2}$--wave
symmetry in
 a $4\times4$ lattice with $n_\up = n_\down = 5$ electrons in the
$t-t'$-Hubbard model and the repulsive interaction $U=4$.}
\end{figure}

Next we study the influence of the $t'$ hopping  for repulsive
interaction $U=4$. Quantum Monte Carlo calculations show a plateau for 
the d$_{x^2-y^2}$ correlation function in the $t-t'$-Hubbard model
\cite{MOR94}, \cite{ZHA97}. 
Figure \ref{posutprime} shows the average 
correlation functions with $d_{x^2-y^2}$ symmetry.

Here the full correlation function is nearly independent of $t'$ for $t'>-0.4$.
The vertex correlation function is much smaller than in the attractive case
(cp.\ fig.\ \ref{negutprime} and fig.\ \ref{posutprime}).
The increase of $\bar{C}_{\rm d}^{\rm v}$ by a factor of about 10 between
$t'=0$ and $t'=-.45$ is much larger than the increase of
$\bar{C}_{\rm s}$ and $\bar{C}_{\rm s}^{\rm v}$ in the attractive $t-t'$-Hubbard
model and of $\bar{C}_{\rm d}$ in the repulsive model.

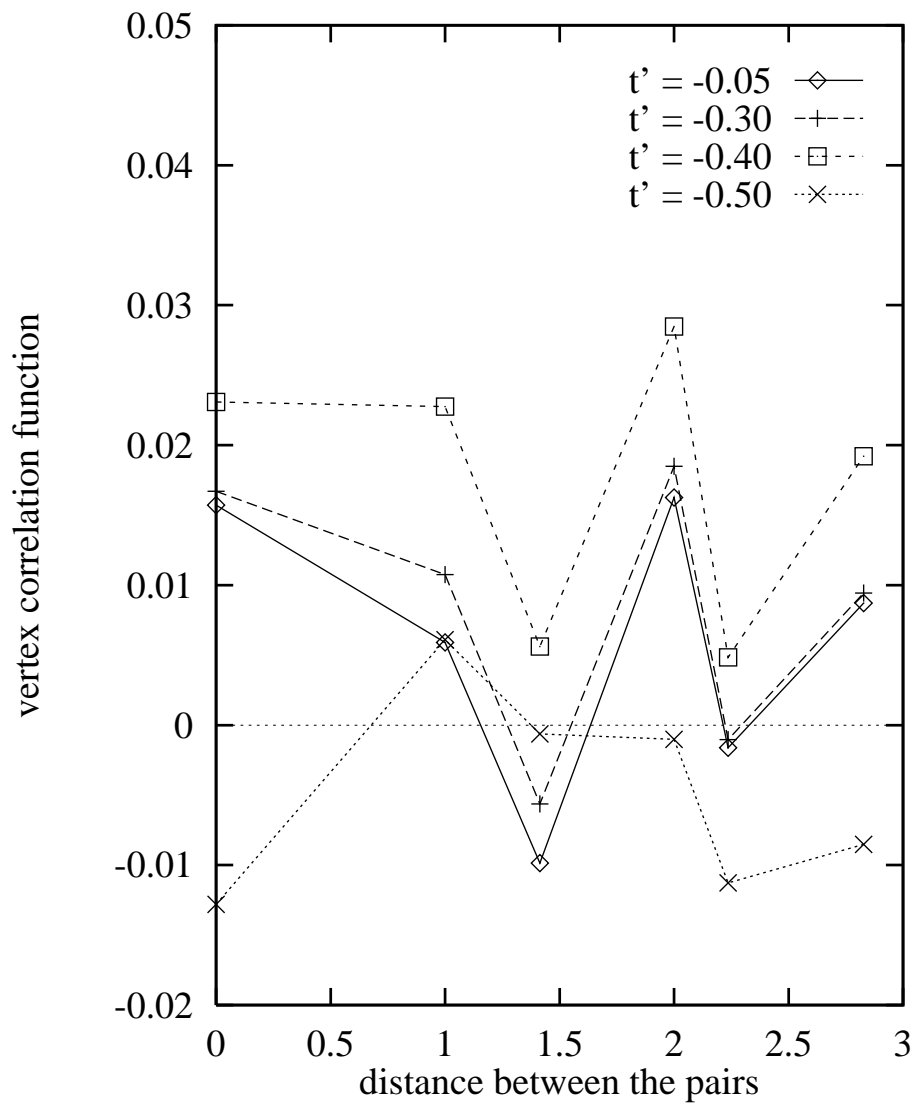
\begin{figure}[p]
\caption{
\label{posulong}
Vertex correlation ($C_{\rm d}^{\rm v}(|r|)$) function 
with d$_{x^2-y^2}$--wave symmetry in a $4\times4$ lattice with 
$n_\up = n_\down = 5$ electrons in the
$t-t'$-Hubbard model and $U=4.0$.} 
\end{figure}

In figure \ref{posulong}
the vertex correlation function with $d_{x^2-y^2}$ symmetry 
is plotted against the distance $|r|$ of the "cooper pairs".
For $t'=-0.40$ the vertex correlation 
function has, in contrast to larger $t'$, no longer a negative value at 
$|r|=\sqrt{2}$ and is positive for all distances $|r|$. 
This negative value at $|r|=\sqrt{2}$ is also seen in the 
quantum Monte Carlo results for larger systems \cite{HUS94}, \cite{HUS97}.

In the case of $t'=-0.50$, where the gap $g(t')$ gets zero, 
$C^{\rm v}_{\rm d}(|r|)$
changes its shape completely and most values are negative 
and also the average is negative. 
In contrast to the attractive $t-t'$-Hubbard model, where 
the plateau is decreasing gradually, in the
repulsive case one observes a complete break down of the plateau for $t' 
= -0.5$.
This means again, that the vertex correlation function depends  
strongly on the energy dispersion
$\varepsilon_k$ (eq.\,\ref{eqepsilon}), which is transformed due to the 
$t'$-hopping. As in the attractive $t-t'$-Hubbard model the maximum of 
the  superconducting correlations is not at $t'=-0.37$, where a 
Van-Hove singularity is in the non interacting infinite system. 

To clarify the origin of this behavior it is necessary to study 
larger systems. For a possible connection with the Van-Hove-scenario
it is also necessary to calculate the density of states for this
parameter regime.

\section{Conclusion}

We have presented an effective algorithm for the implementation of the
exact diagonalization of the
$t-t'$-Hubbard model in momentum-space respresentation.
As method for the exact diagonalization we use the Lanczos algorithm.
We showed a detailed description of the parallel algorithm. 
The speed-up of the code is almost linear for the correlation functions
and is increasing with increasing
size of the Hilbert space.

The key point in our algorithm is a new method of labeling the states 
that is compact and in contrast to previous methods \cite{GAG86}, \cite{LEU92}
also gives a consecutive order without
any interruption. This makes the distribution on the different
processes for parallelization  a straightforward task.
With the access to powerful modern parallel computers we were able 
to scan the parameter space of the $t-t'$-Hubbard model in more detail.

The influence of the $t'$-hopping to the ground state energy 
is vanishing with increasing interaction strength 
in the attractive $t-t'$-Hubbard model.
This is in contradiction to the correlation functions with on site s-wave
symmetry, where the influence of $t'$ is remaining for all studied 
attractive interactions.

The average full and vertex on site s-wave correlation functions 
have a broad maximum at $t'=-0.5$, where the gap $g(t')$ is vanishing,  
in the attractive  $t-t'$-Hubbard model. 

In the repulsive $t-t'$-Hubbard model only the d$_{x^2-y^2}$ vertex
correlation functions show a strong increase with decreasing $t'$ and
gap $g(t')$. At $t'=-0.5$ and $g(t')=0$ $\bar{C}^{\rm v}_{\rm d}$ has
a break down to a negative value. 

The origin of this behavior and a possible connection with the Van-Hove
scenario for high $T_c$ superconductors is yet not clear. Simulations
for larger systems are necessary.

\section{Acknowledgment}

We are grateful for the Leibnitz Rechenzentrum M{\"u}nchen (\,LRZ\,)
for providing us a generous amount of CPU-time on their IBM SP2 parallel
computer.
Werner Fettes wants to thank  the "Deutsche Forschungs Gemeinschaft"
(\,DFG\,) for the financial support.


\bibliographystyle{unsrt}

\newpage


\begin{table}[p]
\caption{
\label{tabstates}
Size of the Hilbert space in a $4\times 4$ system. 
$M_\sigma$: number of states for the electrons with spin $\sigma$,
$M_k^\sigma$: number of states with spin $\sigma$ and momentum $k$, 
$M$: size of the Hilbert space,
$M_t^K$: size of the subspace of the Hilbert space for the states with
momentum $K=(0,0)$.
}
\end{table}


\end{document}